\documentclass[12pt]{article}
\usepackage{epsfig}

\def\beq{\begin{equation}}
\def\eeq#1{\label{#1}\end{equation}}
\def\eeqn{\end{equation}}

\def\beqa{\begin{eqnarray}}
\def\eeqa#1{\label{#1}\end{eqnarray}}
\def\eeqan{\end{eqnarray}}

\let\bar=\overbar

\def\Dslash{\not{\hbox{\kern-4pt $D$}}}
\def\dslash{\not{\hbox{\kern-2pt $\del$}}}

\def\ee{e^+e^-}

\def\msb{{\bar{\ssstyle M \kern -1pt S}}}

\def\Title#1{\begin{center} {\Large {\bf #1} } \end{center}}

\begin{document}

\Title{Total Cross-sections at very high energies: from protons to photons\footnote{Presented at the Tenth Workshop on Non-Perturbative 
Quantum Chromodynamics
at
l'Institut Astrophysique de Paris
June 8-12, 2009}}

\bigskip\bigskip


\begin{raggedright}  

{\it  R.M. Godbole$^1$, A. Grau$^2$, G. Pancheri$^3$ and Y. Srivastava$^4$\index{G. Pancheri}\\
1. Centre for High Energy Physics, Indian Institute of Science, Bangalore,
 560012, India.\\
2. Departamento de F\'\i sica Te\'orica y del Cosmos, Universidad de Granada,
 18071 Granada, Spain. \\
3. INFN Frascati National Laboratories, Via Enrico Fermi 40, I-00044 Frascati,
 Italy.\\
4. INFN and Physics Department, University of Perugia, Via A. Pascoli
, I-06123 Perugia, Italy, \\
and \\
Northeastern University, Boston, Massachussetts 02115, USA
}
\bigskip\bigskip
\end{raggedright}

\section{Introduction}
A good knowledge of total crosss-sections of high energy photon-proton and 
photon-photon interactions, in an energy domain where no data are available,
is important from the point of view of understanding high energy cosmic ray 
data and planning of the TeV energy $\ee$ colliders respectively. This requires
developing a model which can explain the current energy dependence of these
cross-sections observed in the laboratory experiments and then using 
it to predict  the cross-sections in the
required  higher energy regime. Apart from this very prosaic reason for 
studying the subject, high energy behaviour of total hadronic cross-sections 
is an issue of great theoretical importance. Very general arguments based on
unitarity, analyticity and factorisation in fact imply a bound on the high
energy behaviour of total hadronic cross-sections~\cite{froissart}. This 
bound predicts,  independent of the  details of the strong interaction 
dynamics, that asymptotically $\sigma_{tot} \le C (\log {s})^2$. 
All the experimentally measured hadronic cross-sections seem to rise with 
energy~\cite{PDG}, although it is not clear whether the
rate is the same for all the hadronic processes; nor is it clear
whether the asymptotic behaviour is already reached at the current energies.
In view of the important clues to the strong interaction dynamics that
this energy dependence holds and the equally strong need of its precise
knowledge in the high energy regime for the planning of future experiments,
or the understanding of the high energy cosmic ray data, it is not 
surprising that this has been a subject of intense theoretical investigations
~\cite{martinreview,tan}.  

Since QCD has now been established as {\it the} theory of strong interactions,
it is of course of import to seek an understanding of this interesting
 bound in a QCD based picture and try to see clearly as to which piece of 
the QCD dynamics is it most closely related to. A description of one effort
~\cite{ourmodel}
in this direction, viz. a QCD based model to describe the energy dependence 
of the total hadronic cross-section and its extension \cite{EPJC} to the case of the high
energy behaviour of the photon induced processes is the subject of this note.
We will first summarise very briefly the current experimental situation
on the observed energy dependence for  all the hadronic cross-sections 
including the photon induced ones. Then we will describe the original 
minijet model~\cite{minijet} which tries to calculate this dependence in
a QCD based picture. After pointing out the problem of 'too fast' an energy
rise predicted in  these models, we will then show how the resummation of
soft gluons can tame the rise~\cite{froissartus} and how it is possible
to obtain a satisfactory description of the current data with stable predictions
at the LHC~\cite{plb659} within the framework. Then we turn to how the model
can be extended to describe photon-induced processes in terms of the measured
photon structure function and present our predictions for the high energy 
photon-proton cross-sections obtained in this model~\cite{EPJC}.

\section{The data for $pp$ to $\gamma \gamma$}
\begin{figure}[htb]
\begin{center}
\epsfig{file=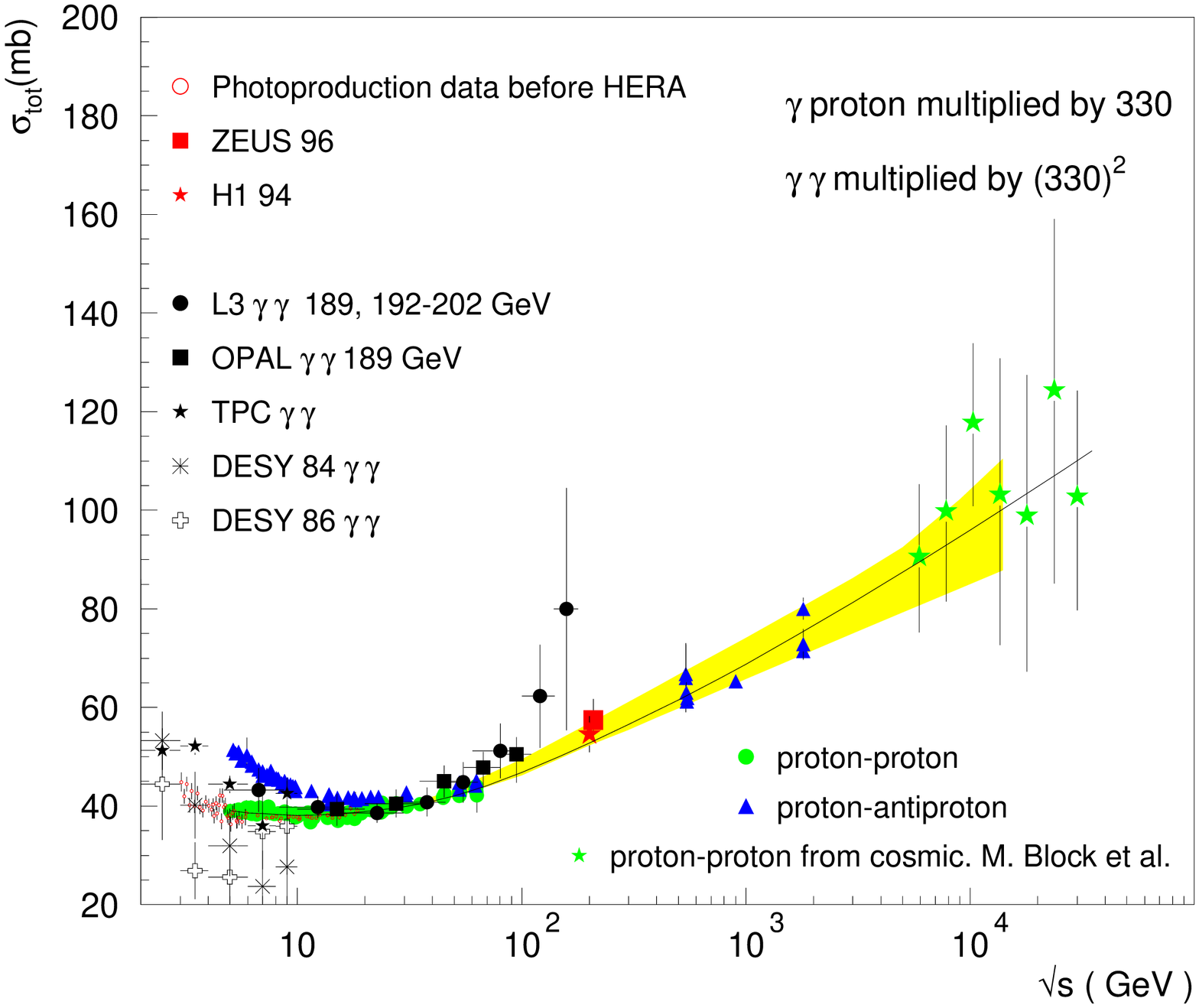,width = 3.0in, height=2.2in}
\hspace{-0.3cm}
\epsfig{file=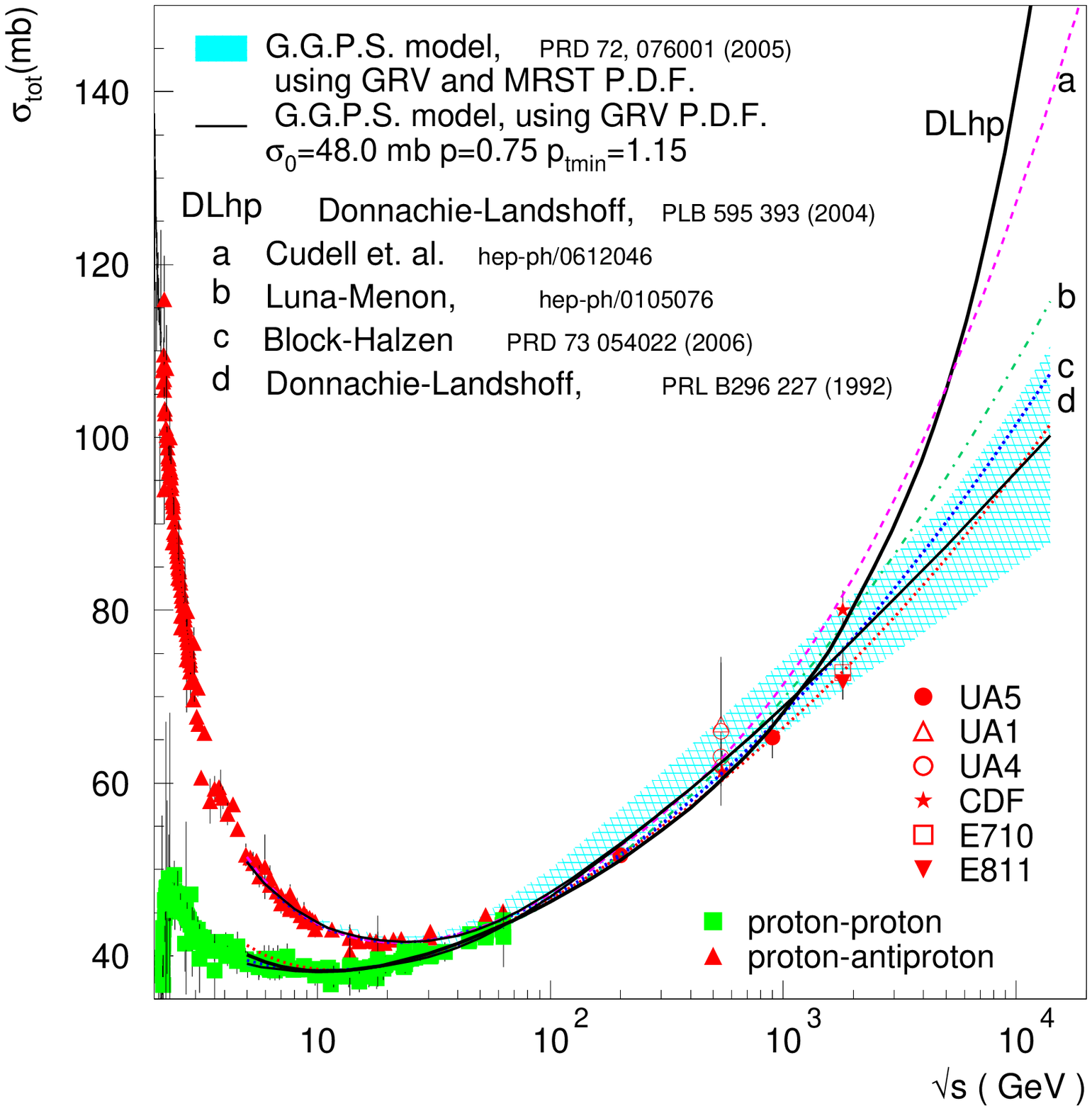,width= 3.0in, height=2.1in}
\caption{Proton and photon normalised cross-sections ~\protect\cite{EPJC} 
with the band of predictions for proton-proton  from \protect~\cite{plb659}, 
also shown on the right, together with   estimates from various models.}
\label{fig:1}
\end{center}
\end{figure}

With the $\bar p p/pp $  
, $\gamma p$ 
and $\gamma \gamma$ 
cross-sections in the milibarn,  microbarn and nanobarn range\cite{comment1}, 
 it is possible to accommodate them all 
in the same figure by multiplying the $\gamma p$ and $\gamma \gamma$ 
cross-sections by $330$ and $(330)^2$ respectively.  This is an {\it ad hoc} 
factor, approximately given by Vector Meson Dominance (VMD) and the quark
 parton model \cite{EPJC}. 
Such compilation of all the proton and photon data for the hadronic 
 cross-sections is shown in Figure~\ref{fig:1},  from~\cite{EPJC}.  
While $\gamma p$ and $\bar p p/pp$ data could  be interpreted as indicating 
the same rise, at least  at presently reached accelerator energies, the figure 
seems to indicate  that the  $\sigma^{\gamma \gamma}$ data from LEP  
rise somewhat faster than the others. 
 In the left panel of Fig.~\ref{fig:1} the band corresponds to the predictions
 for the total $\bar p p/pp$ cross-sections according to the Block-Nordsieck
 (BN) model which we shall describe in the coming section. The same band
is also shown  in the right panel, together with predictions of other 
models~\cite{pp_models} for purely proton processes.    
 The figure  makes clear that  any model has to address three 
issues: 1) what makes the cross-sections rise,  2) what makes them
obey the Froissart bound and  3) whether they indicate any breakdown
 of factorisation between the proton and photon processes. 

\section{Minijet models}
Minijet models were one of the early QCD based models which made an effort to 
understand the energy rise in terms of the rising gluon content of the hadrons
at small values of $x$ and the basic QCD cross-section \cite{levin}.
The cross-section for the  jet production in collisions of two hadrons $A,B$ 
in the process:
\begin{equation}
A + B \rightarrow X + jet  
\end{equation}
is obtained by convoluting  the parton-parton subprocess cross-section with 
the given parton densities and integrating over all values of incoming 
parton momenta and outgoing parton transverse momentum $p_t$, according to 
the expression
\begin{eqnarray}
\sigma^{AB}_{\rm jet} (s,p_{tmin})=
\int_{p_{tmin}}^{\sqrt{s}/2} \!\! d p_t \int_{4
p_t^2/s}^1 \! \!\! d x_1  \int_{4 p_t^2/(x_1 s)}^1 \! \! \! \! d x_2 
 \nonumber \\
\times \sum_{i,j,k,l}
f_{i|A}(x_1,p_t^2) f_{j|B}(x_2,p_t^2)
\frac { d \hat{\sigma}_{ij}^{ kl}(\hat{s})} {d p_t} 
\end{eqnarray}
where $A$ and $B$ are the colliding hadrons or photons. 
By construction, this cross-section  depends on 
the particular parametrization of the 
parton densities evaluated at scale $p_t^2$. 
This cross-section strongly depends on the lowest $p_t$ 
value on which one integrates, viz. $p_{tmin}$. The term 
{\it mini-jet} was introduced 
long ago~\cite{pancheri-rubbia} to indicate all those low $p_t$ processes 
which 
are amenable to a perturbative QCD calculation  but are actually not observed 
as hard jets;  $p_t$ indicating the scale at which $\alpha_s$ is
evaluated in the mini-jet cross-section calculation.  One can have  
$p_{tmin}\approx 1\div 2\ GeV$.
This minijet cross-section  rises very fast with $\sqrt{s}$, the rate of
rise is controlled by the dependence of the parton densities at low-x values 
and  $p_{tmin}$ used.  In the simplest version of the model
 \cite{minijet}
$\sigma^{tot}$ was assumed to be given by

$$
\sigma^{tot} = \sigma^0 + \sigma^{jet} (s,p_{tmin}).
$$
Eventhough these early calculations caught the essence of QCD dynamics
that can cause the rise of total and inelastic cross-section 
with energy, they  predicted a rise with energy which depended on a
parameter $p_{tmin}$ arbitrarily varying with energy. This was made necessary by  the known low-x
behaviour of the parton densities, which predicted a very fast rise with energy,
leading to unitarity violation.  
The unitarity is achieved by embedding the minijet cross-section in an
eikonal picture~\cite{eikonalminjet}. In fact many features of not just the
cross-section but also quantities such as multiplicity distributions etc.
can be successfully described in the eikonalised minijet 
picture~\cite{tjostrand}. 

 In the eikonal formulation, $\sigma^{tot}$ is obtained from an eikonal
 function $\chi(b,s)$ which
describes  the impact parameter distribution during the collision, namely
\begin{eqnarray}
\sigma_{elastic}= \int d^2{\bf b} |
1-e^{
i\chi(b,s)
}
|^2 \\
\sigma_{total}=
2
\int d^2{\bf b} 
[
1-e^{
-{\cal I} m \chi(b,s)
}cos\Re e\chi(b,s)
]\\
\sigma_{total\  inelastic}\equiv \sigma_{inel}=\int d^2{\bf  b} 
[
1-e^{
-2 {\cal I} m \chi(b,s)
}]
\label{sigmainel}
\end{eqnarray}
Neglecting the real part of the eikonal for the hadronic processes, 
one gets a very simple expression for the total cross-section
 \begin{equation}
\label{sigtotsimple}
 \sigma_{total}=
2
\int d^2{\bf b} 
[1-e^{-{\bar n}(b,s)/2}]
\end{equation}
with ${\bar n}(b,s)$ the average number of inelastic collisions.

Minijet models with the well motivated QCD input, embedded into the 
 eikonal representation, are able to describe  the early rise correctly.
However, without any 
further energy dependent input they often fail to obtain the presently 
exhibited  levelling off at high energy, a behaviour already consistent
 with the Froissart bound \cite{martinhalzen}.  It is clear that one needs 
additional input: within QCD there is one important effect which can modify 
the energy dependence, and this is  soft gluon emission from the scattering
 partons. We shall discuss this effect in the next section.  

\section{Block-Nordsiek Model}
 The BN model differs from  and improves on the usual mini-jet models,
 including the eikonalized ones, in three significant ways, namely 1) by
 implementing perturbative QCD input through currently  used PDFs
for the mini-jet cross-sections, 2) introducing soft gluon $k_t$-resummation 
to control the rise as the basic mechanism which  describes  the impact
 parameter distribution of the collision,  with the upper scale in soft gluon 
resummation linked by  kinematics  to the mini-jet cross-section and finally 
3) pushing the soft gluon integral into the InfraRed (IR) region. We have
 discussed  the mini-jet calculation in the previous section. Here we shall
 illustrate the basic features of resummation and our approach to it.
\subsection{Resummation }
We shall start by recalling some properties of soft photon resummation.
 In QED,  the general expression for  soft photon resummation in the 
energy-momentum variable
$ K_\mu$  can be obtained order by order in perturbation theory 
\cite{jr,lomon,yfs} as 
\begin{equation}
d^4P( K)=d^4  K
\int {{d^4x} \over {(2 \pi)^4}}
e^{i{ K \cdot x} -h( x,E)}
\label{d4pk}
\end{equation}
where $d^4P(K)$ is the probability for an overall 4-momentum $K_\mu$ 
escaping detection,  
\begin{equation}
h(x,E)=\int d^3{\bar n}(k)[1-e^{-ik\cdot x}]
\label{hdE}
\end{equation} 
and
\begin{equation}
d^3{\bar n}(k)=
{{d^3k}\over{2k_0}}
|
j_\mu(k,\{p_i\})
|^2
\label{d3n}
\end{equation}
The electromagnetic current $j_\mu$ depends on the momenta of the emitting
 matter fields $\{p_i\}$ and to leading order in $\alpha_{QED}$ is given by
\begin{equation}
j_\mu(k,\{p_i\})=-{
{ie}
\over{
(2\pi)^{3/2}
}}
\sum_i\epsilon_i{{p_{i\mu}}\over{p_i\cdot k}}
\label{jmu}
\end{equation}
with the sum running on all the entering particles ($\epsilon_i=+1$) and
 antiparticles(-1). For outgoing fields, the signs are reversed.

The expression in Eq.( \ref{d4pk}) can also be obtained employing the methods 
of statistical mechanics
\cite{ept}. One starts  with a discrete variable representation 
\begin{equation}
 d^4P(K)= \sum_{n_k} 
  P(
   \{
   n_k \}
   )
  \delta^4(K-\sum_k n_k k)d^4K 
 \label{d4p}
 \end{equation}
and sums on all the possible   Poisson distributed soft photon configurations,
 i.e.
\begin{equation}
P( \{n_k\})=\Pi_{k}\frac{{\bar n}_k^{n_k}}{n_k!}exp[-{\bar n}_k ]
\label{poisson}
\end{equation}
where $n_k$ is the number of photons emitted with momentum ${\bf k}$.
 From Eqs.~(\ref{d4p},\ref{poisson}) one can determine the probability of observing
 a 4-momentum loss K accompanying a charged particle reaction by using the 
integral representation for the 4-dimensional $\delta$-function and 
exchanging the order between forming the product in $P(\{ n_k\})$ 
and the summation over all the distributions.
 Then,  going  from the sum over discrete values to an
 integral over soft photon momenta, one can obtain  Eqs.~(\ref{d4pk},
\ref{hdE}). 
The  derivation being semi-classical, it cannot give any information on 
$d^3{\bar n}(k)$. For this, one needs to use the perturbative expression for 
the electromagnetic current as in Eqs.~(\ref{d3n},\ref{jmu}). Notice that in 
this derivation,  it is energy momentum conservation  which 
ensures   the cancellation of the infrared divergence between soft and real 
quanta emission. This follows from the semi-classical approach, since  in the 
infrared region the uncertainty principle does not allow to distinguish 
between real and virtual quanta.

From Eq.~(\ref{d4p}), the integration over the 3-momentum variable gives
\begin{equation}
\label{dpo}
dP(\omega)={{\beta}\over{\gamma^\beta \Gamma(1+\beta)}}
{{d\omega}\over{\omega}}({{\omega}\over{E}})^\beta 
\end{equation}
with $\gamma$ the Euler's constant,   $\beta \equiv \beta(\{p_i\},m_e)$
 defined as the integration over the soft photon angular distribution, i.e. 
 $\beta$ such that 
\begin{equation}
\int_{\Omega_k} d^3{\bar n}(k)=\beta {{dk}\over{k}}
\end{equation}
 and $E$ is the maximum energy allowed to soft photon emission. With
 $E$ being the energy scale of the reaction, the integration of 
Eq.(\ref{dpo}) up to the energy resolution $\Delta E$, gives the well
 know behaviour $(\Delta E/E)^{\simeq \alpha \log{E/m_e}}$ proposed in the 
early days of QED \cite{schwinger}.

 While it is easy to obtain a closed form for the energy distribution,
a closed form for the momentum distribution is not available.  It is also not necessary, since the first order expression in $\alpha_{QED}$ is adequate.  More interesting, for applications to strong interactions, is the transverse momentum distribution, namely
 \begin{equation}
 d^2P({\bf K_\perp})=d^2{\bf  K_\perp} {{1} \over{(2\pi)^2}}\int
d^2 {\bf b}\ e^{-i{\bf K_\perp\cdot b} -h( b,E)}
\label{d2p}
 \end{equation}
 with
\begin{equation}
 h(b,E)=\int d^3{\bar n}(k) [1-e^{i \bf k_\perp\cdot b}]
 \label{hbe}\end{equation}
For large   transverse momentum values, 
 by neglecting the second term and using a cut-off term as lower limit of 
integration, the above expression coincides with the   Sudakov form factor 
\cite{sudakov}. In QCD, Eq.~(\ref{hbe})
is used  to discuss  low-$p_t$ transverse momentum distributions through
 the expression \cite{ddt,pp}
\begin{equation}
h( b,M) =  \frac{4C_F}{\pi}
\int_{1/b}^M 
  \alpha_s(k_t^2) {{d
 k_t}\over{k_t}}\log{{2M}\over{k_t}} 
 \label{hbq}
\end{equation}
where $C_F=4/3$.

We   use resummation to approach the very  large impact parameter region, 
which plays a role in such quantities as the total cross-section. For this, 
we need to explore the IR region and use the full range of integration of
 Eq.~(\ref{hbe}). In the region $k_t<\Lambda$ we propose to use an 
expression for the soft gluon spectrum which takes into account the
 effect of a rising confining potential. The expression to use will then be 
singular as $k_t\rightarrow 0$, but must be integrable. As discussed in many 
of our publications \cite{ourmodel,froissartus}, we shall interpolate between the
 asymptotic freedom region (AF) and the IR through the following expression 
for the strong coupling constant
\begin{equation}
\label{alphapheno}
\alpha_s(k_t^2)={{12 \pi }\over{(33-2N_f)}}{{p}\over{\ln[1+p({{k_t^2}
\over{\Lambda^2}})^{p}]}}
\end{equation}
which coincides with the usual one-loop formula  for
 values of $k_t>>\Lambda$, while going to a singular
 limit for small $k_t$,  and 
generalizes Richardson's ansatz for the one gluon exchange potential  to
 values of $p\le 1$ \cite{richardson}. However, notice that,  in order to 
make the integral finite, we
 need  $p < 1$. Also analyticity requires $p>1/2$ \cite{froissartus}.

Having thus extended the region of interest in Eq.~(\ref{hbq}) to the large
 impact parameter values, we use the Fourier transform of Eq.~(\ref{d2p})
 to describe the impact parameter distribution, and input it into the average
 number of hadronic collisions, ${\bar n}(b,s)$, generated by mini-jets, 
i.e. we write 
\begin{equation}
 n_{hard}(b,s) =A_{BN}(b,s)\sigma_{jet}
\end{equation}
with
\begin{equation}
\label{ourAB}
A_{BN}(b,s)\equiv A_{BN}(b,M)=A_0 e^{-h(b,M)}
\end{equation}
with the normalization constant $A_0$  
\begin{equation}
A_0=\frac{1}{2\pi \int bdb \ e^{-h(b,M)}}
\end{equation}
and with the integral in Eq.~(\ref{hbq}) extended down to $k_t=0$, 
to be used in  Eq.~(\ref{sigtotsimple}). The subscript BN indicates that this 
$b$-distribution 
is obtained from soft gluon $k_t$-resummation into the IR region.

Another  important quantity in soft gluon resummation,  is the upper limit of 
integration, the scale $E$ in QED calculation, and which we have indicated 
with $M$ in the QCD integral. $M$ is in general a function of the incoming 
and 
outgoing parton
 momenta, and can be determined by the kinematics, as discussed  in 
\cite{EPJC} and following \cite{mario}. Upon integration over all the PDFs, 
it is seen to be of the order of $p_{tmin}$ and slowly varying with the c.m.
 energy.

\subsection{BN model for protons}
In order to apply our QCD description, inclusive of mini-jets and soft 
$k_t$-resummation, to scattering processes, we split the average number of 
collision ${\bar n}(b,s)$ as 
\begin{equation}
{\bar n}(b,s)=n_{NP}(b,s)+n_{hard}(b,s)
\end{equation}
and parametrize the first term, which is meant   to include all 
those processes which cannot be described through parton-parton scattering 
discussed in the previous section. At low energies, i.e.
$\sqrt{s}$ approximately up to $20\div 30\ GeV$, ${\bar n}(b,s)
\approx n_{NP}(b,s)$,
whereas as the c.m. energy increases, the second term will asymptotically
 dominate. To see how, in the BN model, soft gluon resummation helps to tame
 the rise of the mini-jet cross-section and bring satisfaction of the Froissart
 bound, we shall thus look at the expression for $\sigma_{total}$ at
 extremely high energies, namely
\begin{equation}
\label{sigmaT}
\sigma_T(s)=2\int d^2{\bf b}[1-e^{-n_{hard}/2}]
\end{equation}
In the previous edition of this Conference, we have presented a preliminary 
version of the argument for satisfaction of the Froissart bound in our model
\cite{paris07}. A more accurate argument has now been   given in 
\cite{froissartus}. Basically, we find that extending the integration in 
Eq.~(\ref{hbq}) into the IR region with our IR singular, but integrable,
 $\alpha_s$ introduces a  cut-off in $b$-space, which behaves 
at least like an exponential ($1<2p<2$). Such cut-off  
 allows to evaluate the integral of Eq.~(\ref{sigmaT})
at the value where the integrand suddenly decreases. Inserting  the
 asymptotic expression for $\sigma_{jet}$ at high
energies, which grows like  a power of $s$, 
and  $A_{BN}(b,s)$ from Eq.~(\ref{ourAB}),
in such large-$b$, large-$s$ limit,
 we  obtain 
\begin{equation}
  n_{hard}  = 2C(s)e^{ - (b{\bar \Lambda})^{2p} }
\end{equation}
where  $2C(s) = A_0(s) \sigma _1 (s/s_0 )^\varepsilon  $. The resulting
expression for $\sigma_T$ 
\begin{equation}
  \sigma _T (s) \approx 2\pi \int_0^\infty  {db^2 } [1 - e^{ -
C(s)e^{ - (b{\bar \Lambda})^{2p} } } ]
\label{sigT}
\end{equation} 
leads to
\begin{equation}
{\bar \Lambda}^2\sigma_{T}(s)\approx (\frac {2\pi} {p})\int_0^{u_0} 
du u^{\frac{1-p} {p}}
=2\pi u_0^{1/p}
\end{equation}
with
\begin{equation}
u_0=\ln[\frac {C(s)}{\ln 2}]\approx \varepsilon \ln s
\end{equation}
To leading terms in $\ln s$, 
 we therefore derive the asymptotic energy dependence 
\begin{equation}
  \sigma _T  \to [\varepsilon \ln (s)]^{(1/p)}
 \label{froissart}
\end{equation}
 Remembering that $1/2<p<1$ \cite{froissart},  the above result shows that, 
with soft  gluon momenta  integrated into the IR region, $k_t<\Lambda$,   and
 a singular but integrable coupling to the quark current, our model leads to 
a behaviour consistent with  the Froissart-Martin bound~\cite{froissart}.
\subsection{The BN model for photons}
To apply the BN model to photon processes, we follow 
refs.~\cite{sarcevic,fletcher}, and  estimate the total cross-section as
\begin{equation}
\sigma_{tot}^{\gamma p}=2P_{had}\int d^2{\bf b} [1-e^{-n^{\gamma p}(b,s)/2}]\ ,  
\ \ \ \ \ \ \ n^{\gamma p}(b,s)=n_{NP}^{\gamma p}+
A_{BN}^{\gamma p}(b,s)\frac{\sigma_{jet}^{\gamma p}}{P_{had}}
\end{equation}
with $P_{had}=1/240\approx {\cal O}(\alpha)$ to represent the probability 
that a photon behaves like
 a hadron and with photon PDF's used to calculate the mini-jet 
cross-sections and the scale energy parameter $M$. A full discussion of how 
to evaluate this parameter ,
called $q_{max}$ when it is averaged over the densities,  can be found in \cite{EPJC}. $q_{max}$
 depends on the energy of the subprocesses and, being evaluated using the PDFs  
of the processes under consideration,  depends
on the specific choice of
the parametrisation used for the  parton densities in the photon and the proton.
As $q_{max}$ increases with energy,  the growth  of the total
 cross-section due 
to  mini jets  is tempered by soft gluon emission. The calculated values of 
$q_{max}$, for all the available parton densities reach some sort of 
saturation 
at high energies, which in turn reflects in the total cross-sections 
reaching a stable slope.

\begin{figure}[htb!]
  \begin{center}
\epsfig{file=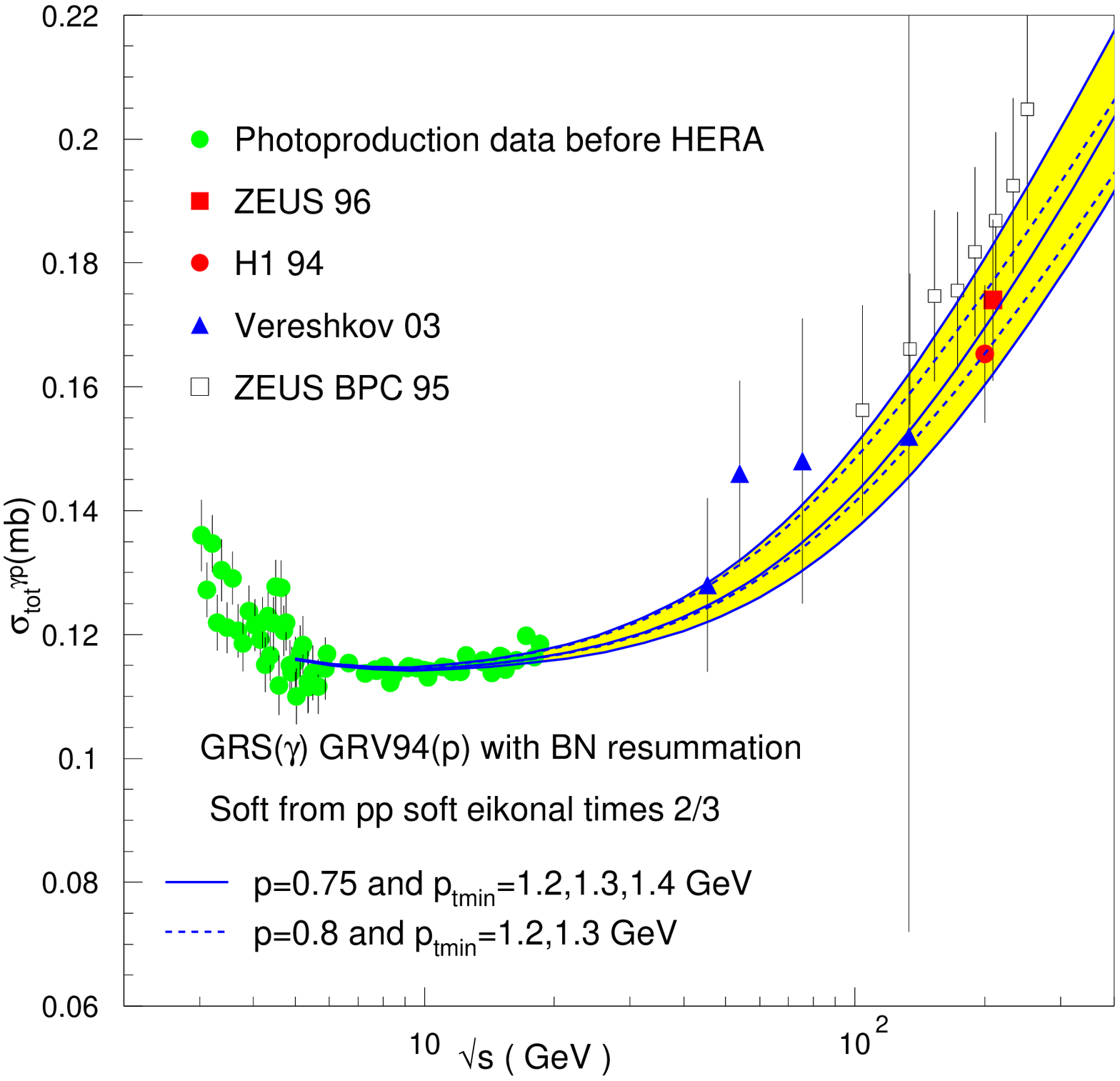,width=3.1in,height=2.5in}
 \hspace{-0.9cm}
 \epsfig{file=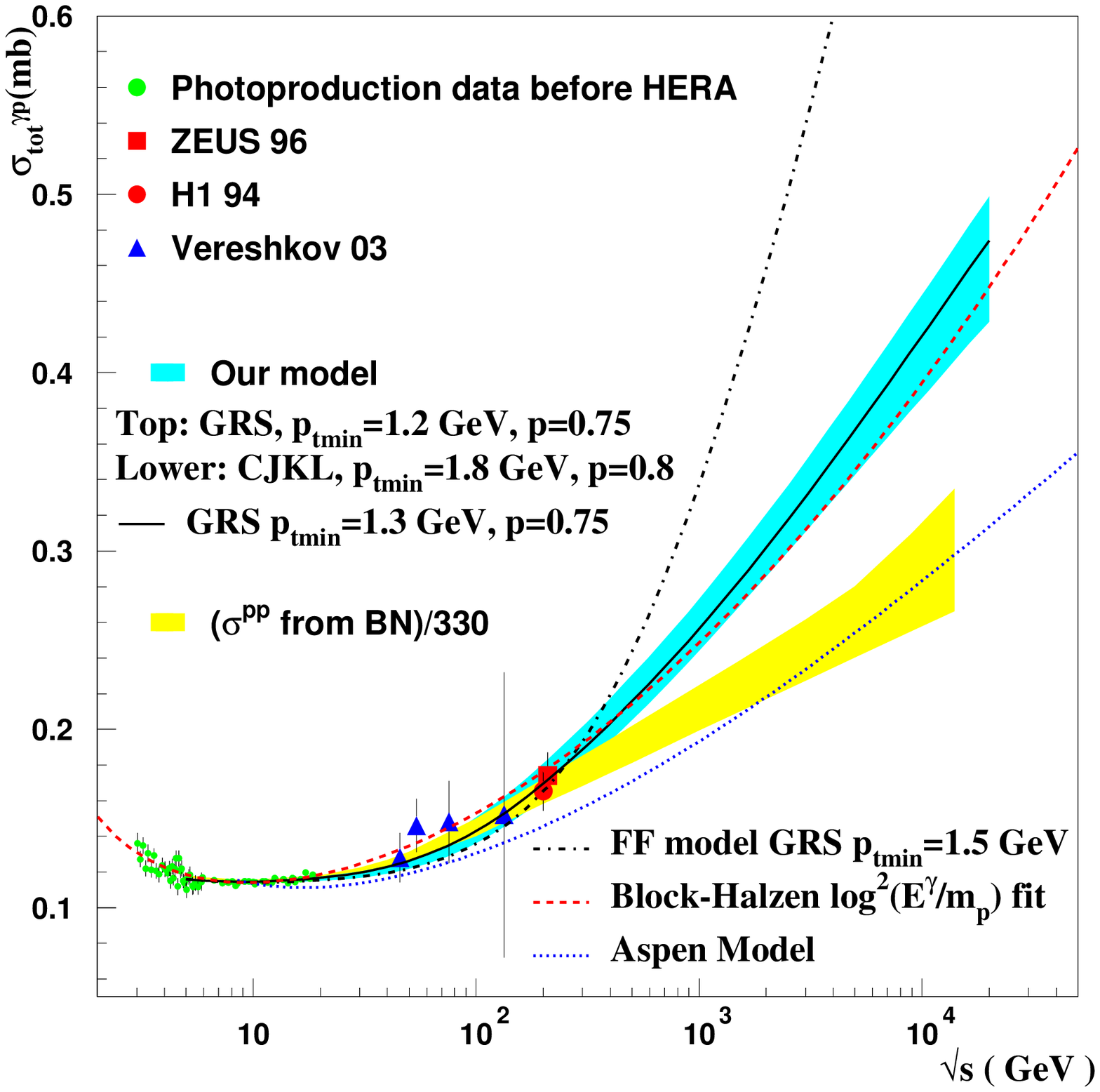,width=3in,height=2.4in}
  \caption{
  $\gamma p$\ total cross section  from \cite{EPJC} in 
different energy ranges.}
  \label{fig:3}
  \end{center}
  \end{figure}

 The application of our model to photon total cross-sections shows some 
interesting features. While in the present accelerator energy range, $\gamma p$
 data could be described also through  factorization models, at very high
 energies in the TeV range, predictions differ. We show this in 
Fig.~(\ref{fig:3}). 
The left panel plots the data up to the highest accelerator energies. Data 
in this region come from cosmic rays \cite{vereshkov}, from extrapolation
 of virtual photon data taken with the BPC \cite{BPC} and from H1 and Zeus 
experiments \cite{datagp}.  In this energy range data can be accomodated by 
many models, including factorization models, in which models for proton 
cross-sections are  extended to photons,  by just multiplying the 
proton curve by a constant factor \cite{bsw} or by extrapolating eikonal
 models with scaling factors in the impact parameter description of photons 
\cite{aspen} or by assuming for photons the same  rise with energy of proton
 cross-section  \cite{dl}. However, as one can see from the right panel, the 
situation changes as the c.m. energy of the $\gamma p$ system reaches into 
the TeV range. Here the discrepancy with factorization models, exemplified 
by the lower band, is clearly indicated both by the upper band obtained 
through our BN model, as well as by the dashed curve within it. This curve 
was independently obtained in \cite{martinhalzen} from a fit to accelerator 
data
and confirms the validity of our model into an energy range so far 
inaccessible through particle  accelerators.

\section{Conclusions}
We have discussed the results from a mini-jet model which incorporates soft 
gluon $k_t$-resummation as a taming effect on the rapid rise with energy 
of low-$x$ initiated mini-jet cross-sections. This has been applied to
both $pp/{\bar p}p$ and $\gamma p$ processes. We find that 
soft $k_t$-resummation, inclusive of IR gluons with  
 $k_t<\Lambda$, plays a crucial role in trasforming the power like rise of the 
jet cross-sections into a more subdued logarithmic behaviour. We accomplish 
this through  the use of a phenomenological ansatz for  the coupling 
between soft gluons and the quark current which gives an expression  
 singular but integrable.
\section*{Acknowledgments}
G.P. thanks for hospitality the Physics Department of Brown University and 
the MIT Center for Theoretical Physics.  R.G. acknowledges support from 
Department of Science and Technology, India for financial support
under Grant No. SR/S2/JCB-64/2007 (J.C. Bose fellowship). 
This work has been partially supported by MEC (FPA2006-\-05294) and  Junta
de Andaluc\'\i a (FQM 101 and FQM 437).


\begin{thebibliography}{99}
\bibitem{comment1}
Our Refs.~\protect\cite{EPJC,plb659} contains the references to all the data plotted
in Figure~\ref{fig:1}.
\end{thebibliography}
\end{document}